\documentclass[12pt]{iopart}

\usepackage{graphicx}
\usepackage{dcolumn}
\usepackage{bm}
\usepackage{epsfig}
\usepackage{stmaryrd}
\usepackage{iopams}

\newcommand{\sub}[2]{{#1}_{ \mbox{\scriptsize #2}}}

\def\beq{\begin{equation}}
\def\eeq{\end{equation}}

\newcommand{\rref}[1]{Ref.~\cite{#1}}
\newcommand{\frefp}[2]{Fig.~\ref{#1}(#2)}
\newcommand{\bref}[1]{(\ref{#1})}

\begin{document}

\title{Supersonic optical tunnels for Bose-Einstein condensates}
\author{S. W\"uster$^{1,2}$ and B. J. D\c{a}browska-W{\"u}ster$^{1,3}$}
\address{$^1$ ARC Centre of Excellence for Quantum-Atom Optics}
\address{$^2$ Department of Physics, Australian National University, Canberra
ACT 0200, Australia}
\address{$^3$ Nonlinear Physics Centre, Research School of Physical Sciences and Engineering, Australian National University, Canberra
ACT 0200, Australia}
\ead{sebastian.wuester@anu.edu.au}

\begin{abstract}
We propose a method for the stabilisation of a stack of parallel vortex rings in a Bose-Einstein condensate. The method makes use of a ``hollow'' laser beam containing an optical vortex. Using realistic experimental parameters we demonstrate numerically that our method can stabilise up to $9$ vortex rings. Furthermore we point out that the condensate flow through the tunnel formed by the core of the optical vortex can be made supersonic by inserting a laser-generated hump potential. We show that long-living immobile condensate solitons generated in the tunnel exhibit sonic horizons. Finally, we discuss prospects of using these solitons for analogue gravity experiments. 
\end{abstract}



\section[Introduction]{Introduction\label{introduction}}
Ultracold quantum gases - Bose-Einstein condensates (BECs) - provide a uniquely accessible testbed for an astonishing range of effects occuring in diverse physical systems ranging from antiferromagnetic materials \cite{damski:kagomepra} and nonlinear photonic structures \cite{elena:nlo} to astrophysical black holes \cite{unruh:bholes,visser:review}. The recent observation of a quantum three-body bound state in an ultracold cesium gas \cite{grimm:efimov} is only one example of a fundamental effect that was predicted decades ago, but had not lent itself to experimental observation in any other physical system. The detailed exploration of these distinct physical phenomena in condensates is enabled by well-developed experimental techniques that provide unprecedented control over the interatomic interactions and the potentials confining the atoms. Advanced trapping schemes have recently been demonstrated by constructing box-like traps \cite{raizen:box} and circular waveguides \cite{gupta:ringtrap}.

\begin{figure}[!ht]
\centering
\epsfig{file={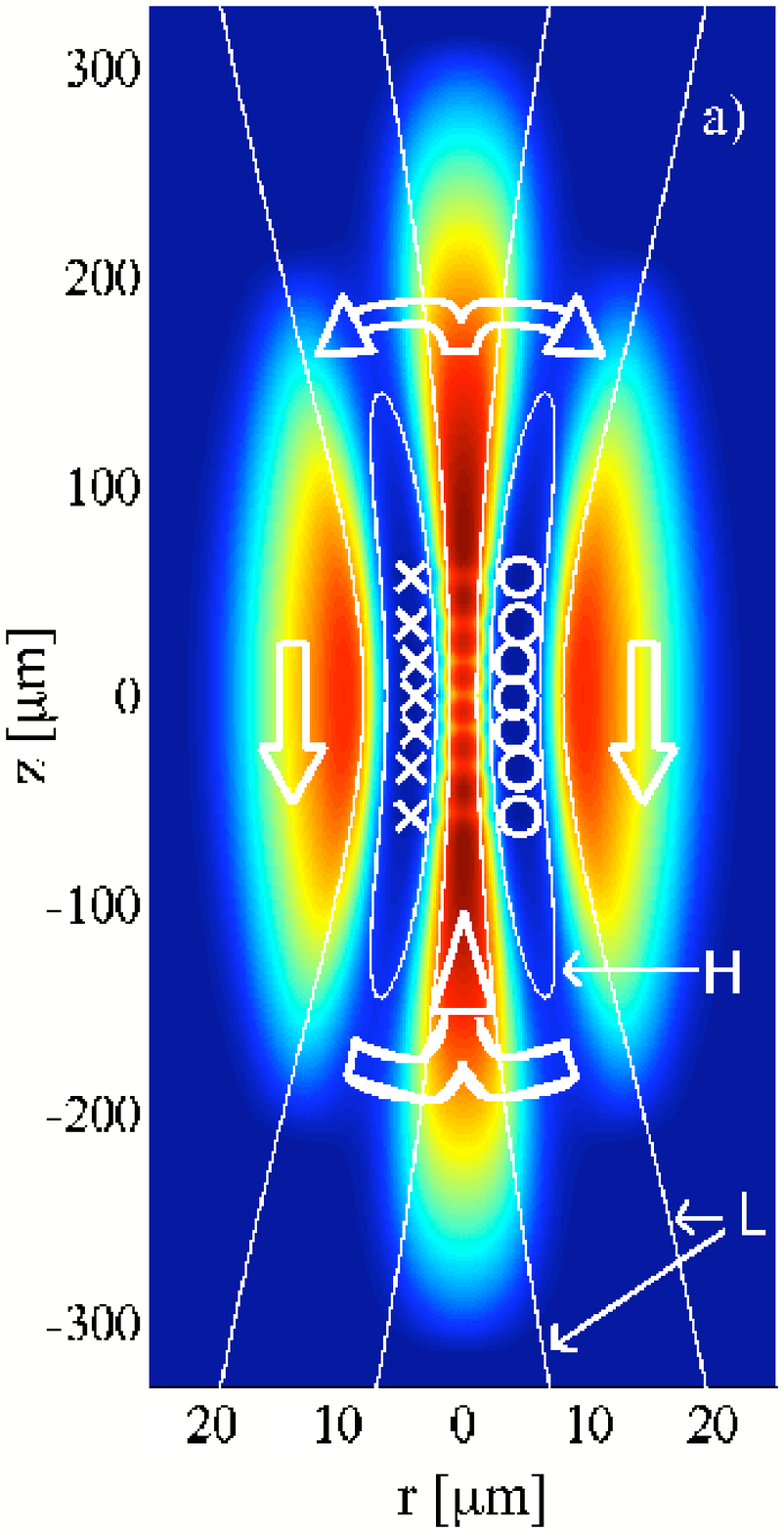},width=5cm} 
\epsfig{file={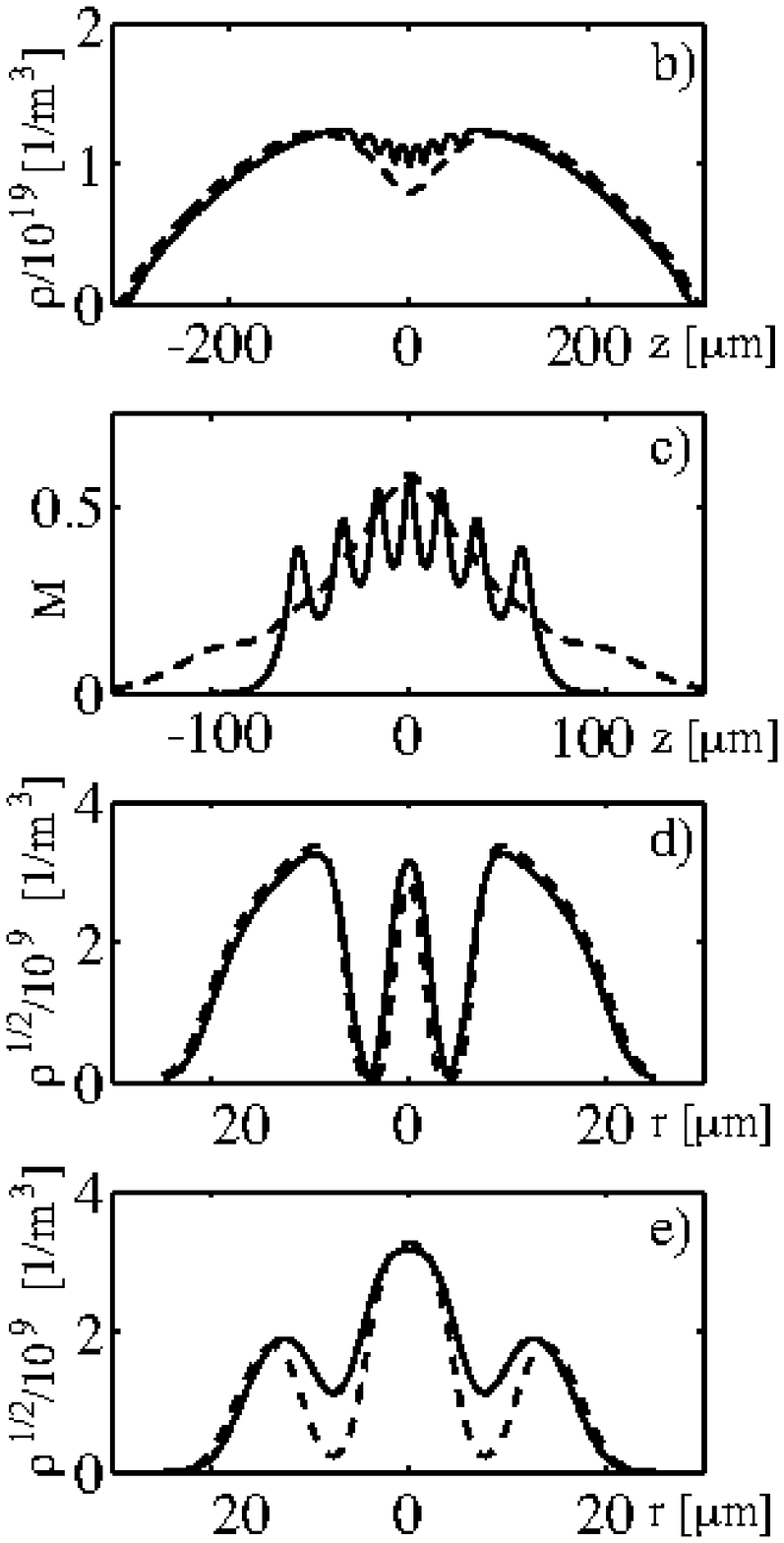},width=5cm}
\caption{Cross-sections of density and velocity structures for different stationary ring flows. 
(a) Shows the BEC density (red highest, blue lowest) in the $rz$ plane for case (i), defined by parameters: $V_{0}=48\hbar\omega_{r}$, $w_{0}=3\mu$m, $q=7$. The arrows indicate the BEC flow, thin white lines are isocontours of the optical vortex potential. 
The low isocontour (marked by L) corresponds to $V_{v}=1 \times 10^{-32}$J, the high one (marked H) to $V_{v}=1 \times 10^{-31}$J. The intersection of ring singularities with the $rz$ plane is marked by $\times$ ($\circ$) for mathematically positive (negative) circulation. 
\\
(b) BEC density and (c) Mach number $M$ at $r=0$ along the $z$ axis. (d) BEC amplitude in radial direction at $z=0$. (e) BEC amplitude in radial direction at $z=170\mu$m. Results for case (i) are indicated by solid lines in (b-e). Results for case (ii): $V_{0}=99\hbar\omega_{r}$, $w_{0}=2.8\mu$m,  $q=9$ are indicated by dashed lines in (b-e). 

\label{2dpicture}}
\end{figure}

In this article we study a Bose-Einstein condensate with repulsive interatomic interactions confined in a harmonic trap and additionally exposed to a laser beam containing an optical vortex~\cite{opticalvortices:review}. By tightly focussing the laser beam blue-detuned from an atomic resonance, it is possible to create a finite length {\it optical tunnel} for the BEC, defined by the ``dark'' region of the beam in the vortex core. This dark region is located between the central vertical white lines in \frefp{2dpicture}{a} and 
corresponds to the potential valley between the potential peaks in \frefp{potentialplot}{a}.

The main purpose of this paper is to show that the proposed trapping configuration can enable the controlled study of a complex topological structure consisting of multiple stationary vortex rings in a BEC. Unsupported ring vortices are energetically unstable against contraction \cite{roberts:ringvortices} and have so far been only observed in non-stationary situations, for example as the decay product of dark solitons or in soliton collisions~\cite{anderson:solitondecay,ginsberg:hybrids}. In contrast, our method could enable the stabilisation of a stack of up to nine parallel vortex rings under realistic experimental conditions. The creation of stationary vortex rings would also aid studies of their interactions \cite{ruost:enginvr} and represent a stepping-stone towards the creation of even more complex matter-wave structures such as skyrmions \cite{ruoste:imprint2,savage:skyrm,wuester:skyrm,battye:homog}. 

A skyrmion is a topologial soliton in a two component BEC, in which the contraction of the ring vortex in one hyperfine component is prevented by filling it with a second hyperfine state carrying a line vortex. This also represents a tool to stabilise multiple ring vortices \cite{ruoste:highw}. 
The optical tunnel system presented here results from a skyrmion if one replaces the hyperfine component carrying a line vortex by an optical field. Our system can thus be thought of as ``atom-light-skyrmion''.

Furthermore, in the proposed trapping configuration the BEC is flowing through the optical tunnel with velocities close to the speed of sound. Therefore we discuss the prospects of using our system for studies of analogue gravity effects. It has been discovered by W. Unruh~\cite{unruh:bholes} that within the hydrodynamic approximation~\cite{book:pethik}, the equations of motion for quantised sound waves in a flowing BEC are analogous to those of a scalar quantum field in curved space time. Within this analogy, a sonic horizon, defined as a surface on which the normal component of the fluid flow becomes supersonic, corresponds to the event horizon of a black hole. Unruh proposed that a sonic horizon created in a BEC flow could enable the experimental observation of Hawking radiation~\cite{hawking:hr1,hawking:hr2} of sound waves in a BEC. 

Here, we show that the BEC flow in the optical tunnel can be driven into the supersonic regime by inserting a laser generated ``hump'' potential into the tunnel. The perturbation can be engineered to result in the development of a single grey soliton in the condensate flow, which is immobile in the lab-frame. The BEC flow past the soliton is supersonic. 

The rest of this paper is organised as follows: In \sref{model_gpe} we introduce our model and describe the optical vortex potential. In \sref{potential} we explain how the total potential has to be adjusted to allow stabilised ring flows, while we present the initial state of our imaginary time method in \sref{imagtime}. In \sref{stationary} we show some exemplary stable ring flows found numerically, followed by a brief description of a scheme for their experimental realisation in \sref{experimental}.
Section \ref{supersonic} contains a discussion of the dynamics that arise, when the flow is driven into the supersonic regime. Finally, we devote \sref{analogue} to an analysis of supersonic flow past a grey soliton in the context of analogue gravity.

\section{Model\label{model}}

\subsection{Gross-Pitaevskii equation\label{model_gpe}} 

We describe the dynamics of the BEC wavefunction $\psi$ with the Gross-Pitaevskii equation (GPE) \cite{stringari:review}:
\begin{eqnarray}
i \hbar  \frac{\partial \psi}{\partial t}&=\left[-\frac{\hbar^2}{2m}\Delta +V(\mathbf{x})
+U |\psi |^{2} \right]  \psi.
\label{3dgpe}
\end{eqnarray}
Using cylindrical coordinates $r$, $z$, $\varphi$, we model a BEC cloud in a cigar-shaped trap with frequencies $\omega_{r}$ radially and $\omega_{z}$ longitudinally. Throughout the paper we assume that $\psi$ does not depend on $\varphi$, hence $\Delta=\frac{\partial^{2} }{\partial r^{2}} + \frac{1}{r}\frac{\partial}{\partial r} + \frac{\partial^{2} }{\partial z^{2}}$. The interaction coefficient $U$ is related to the s-wave scattering length $a_{s}$: $U=4\pi \hbar^{2} a_{s}/m$, where $m$ is the atomic mass. Both parameters are taken for $^{87}$Rb: $m=1.44\times10^{-25}$kg and $a_{s}=5.67$nm. If the wavefunction is written in the polar form $\psi=\sqrt{\rho}\exp{(i \vartheta)}$, we can extract the atom number density $\rho$ and velocity $\mathbf{v}=\hbar\mathbf{\nabla}\vartheta/m$. For later reference we also define the flow speed $v=|\mathbf{v}|$, the local speed of sound $c=\sqrt{U \rho/m}$, the Mach number $M=v/c$ and the healing length $\xi=\hbar/(\sqrt{2}mc)$. Finally, $V=V_{t}+V_{v}+V_{h}$ denotes the external potential whose individual components are:
\begin{eqnarray}
V_{t}(r,z)&=\frac{1}{2}m(\omega_{r}^{2}r^{2} + \omega_{z}^{2}z^{2}),
\label{vfirst}
 \\
V_{v}(r,z)&=V_{0}  r^{2l} \sigma_{0}(z)^{-(l+1)}w_{0}^2  \exp{   \left(   -  r^{2}/\sigma_{0}(z)^{2}   \right)  },
\label{vvort}
 \\
 V_{h}(r,z)&=\sum_{i=1,2}V_{hi} \sigma_{i}(z)^{-2}w_{i}^2  \exp{   \left(   -  r^{2}/\sigma_{i}(z)^{2}   \right)},
 \label{Vhump}
\\
\sigma_{i}(z)^{2}&=w_{i}^{2} +(z-\sub{z}{f}^{(i)})^{2}/(k_{i}^{2}w_{i}^{2}).
\label{vlast}
\end{eqnarray}
The repulsive potential $V_{v}$ can be generated by a blue-detuned optical vortex laser beam~\cite{leach:vortexknots}. The charge of the optical vortex $l$ is taken to be equal 2. In the dynamical simulations we employ a ``hump'' potential $V_{h}$, generated by one or more additional laser beams without any vorticity.
They will be used to create small scale modifications of the overall potential experienced by the atoms within the core of the optical vortex.
The propagation direction of the laser beams coincides with the long trap axis $z$.
Further, $w_{0,1,2}$ are parameters describing the waist of the laser beams at their focus, $k_{0,1,2}$ are the wave numbers of the light and the $\sub{z}{f}^{(i)}$ control the position of the beam focus. For the results presented here we used $k_{i}=1.4\times~10^{7}\mbox{m}^{-1}$ and $\sub{z}{f}^{(0)}=0$.

\subsection{Potential design and trial states\label{potential}} 
\begin{figure}[ht]
\centering
\epsfig{file={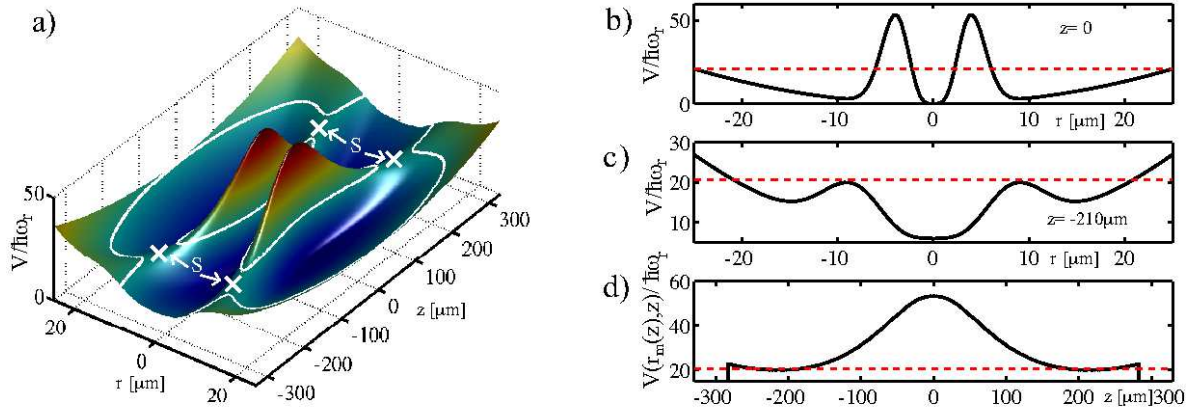},width=\columnwidth} 
\caption{(a) Combined potential $V$ of the harmonic trap $V_{t}$ and optical vortex laser beam $V_{v}$ for $V_{0}=99\hbar\omega_{r}$, $w_{0}=2.8\mu$m,  $q=9$. Trap frequencies: $\omega_{r}=7.8\times 2 \pi$~rad/s and  $\omega_{z}=0.5\times 2 \pi$~rad/s. Saddle points of the potential are indicated by $S$. The white line is an isocontour $V=\mu_{0}$ at $\mu_{0}/(\hbar \omega_{r})=20.9$. (b, c) Radial cross sections of the potential at $z=0$ (b) and at $z=-210\mu$m (c). Dotted lines indicate $\mu_{0}$.  (d) Dependence of the height of the potential ``ridge'' (local maximum) of $V$ on $z$. This corresponds to the function $V(r_{m}(z),z)$ as described in the text. For $|z|>280\mu$m the local maximum disappears for the parameters chosen. 
\label{potentialplot}}
\end{figure}

\Fref{potentialplot} shows a slice in the $rz$ plane of the combined potential $V$ of the harmonic trap and optical vortex $V=V_{t} + V_{v}$.  An equal energy contour is indicated by the white lines. In three dimensions the tight focus of the laser beam generates a circular local maximum, whose intersect with the $rz$ plane yields the peaks situated symmetrically around the origin at $|r|\sim5\mu$m, see \fref{potentialplot}.  
For a suitable choice of parameters the combined potential exhibits saddle points, marked S in \fref{potentialplot}(a). If the value of the chemical potential is above the potential value at the saddle points, the BEC can classically ``wrap'' around the intense region of the optical vortex. 

We find parameter sets for which the saddle points exist using the following method: First we determine the position of the maximum of the vortex potential $V_{v}$ Eq.~\eref{vvort} with respect to $r$, keeping $z$ fixed. We thus solve 
\begin{eqnarray}
\frac{\partial}{\partial r }V_{v}(r,z)&=0
\label{find_extrema}
\end{eqnarray}
for $r$. We ignore the harmonic trap in this step. The solution of Eq.~\eref{find_extrema} is:
\begin{eqnarray}
r_{m}(z)&=\sqrt{l}\sqrt{w_{0}^{2} + \frac{z^{2}}{k_{0}^{2}w_{0}^{2}}}.
\label{maxradius}
\end{eqnarray}
Now taking $V_{t}$ into account, the height of the potential ``ridge'' visible in \fref{potentialplot}(a) is approximately given by $V(r_{m}(z),z)$, where $V_{t}$ is now taken into account. \Fref{potentialplot} (d) shows $V(r_{m}(z),z)$. If saddle points exist, they must be minima of this function. For $V(r_{m}(z),z)$ to have extrema besides the obvious maximum at $z=0$, the quartic expression
\begin{eqnarray}
\fl
\left(\frac{\partial}{\partial z }V(r_{m}(z),z) \right)/z &=
-\frac{2 e^{-l}l^{l} V_{0}w_{0}^{4}}{k_{0}^{2}}\left(w_{0}^{4} + \frac{z^2}{k_{0}^2} \right)^{-2} + \frac{l m \omega_{r}^{2}}{k_{0}^{2}w_{0}^{2}} +m \omega_z^{2}=0
\label{minimumcondition}
\end{eqnarray}
must have real roots. After selecting the optical vortex parameters and $\omega_{r}$, equation \eref{minimumcondition} yields a useful approximation for the strongest axial confinement $\omega_{z}$ allowed.

\subsection{Energetically stable flows\label{imagtime}} 

In order to find energetically stable ring flows we solve Eq.~\eref{3dgpe} for $V=V_{t} + V_{v}$ in imaginary time~\cite{garcia:numerics} and obtain stationary states of the system. The initial trial function for the imaginary time algorithm must have the same topological structure as the target ring flow. To this end, we multiply an amplitude $\sqrt{\rho(r,z)}$ obtained from the Thomas-Fermi approximation~\cite{book:pethik} with a phase factor $\exp{(i\eta(r,z))}$:
\begin{eqnarray}
\psi(r,z,\phi)&=\sqrt{\rho(r,z)}\exp{(i\eta(r,z))}.
\label{seedingstate}
\end{eqnarray}
The phase is given by
\begin{eqnarray}
\eta(r,z)&=q\arccos{\left(\ \frac{|z|}{z} \frac{(D-r)}{s(r,z)}\right)}.
\label{seedingstate_phase}
\end{eqnarray}
The quantity 
\begin{eqnarray}
s(r,z)&=\sqrt{z^{2}+ (D-r)^{2}}
\label{seedingstate_s}
\end{eqnarray}
measures the distance to the single circular ring singularity implicit in Eq.~\eref{seedingstate_phase}.
On any closed loop threaded through this singularity, the phase $\eta$ varies by $2\pi q$, where $q$ is the ring vortex charge. The singularity is located at the point $r=D$, $z=0$ in our coordinates.
An energetically stable excited state with high-charge ring flow corresponds to a local minimum of energy. If the trial condensate function is chosen to have the same topology as the energy minimum, we obtain the stabilised flow after a sufficiently long imaginary time~\cite{garcia:numerics}.
 
\section{Results\label{results}}
\subsection{Stationary states \label{stationary}}
The low density region of a vortex ring in a BEC has a toroidal shape, which matches the shape of the high intensity region of a focussed optical vortex laser beam. In the presence of the optical vortex potential it becomes energetically favourable for the BEC ring singularity to be located at the maximum of laser light intensity. This prevents the contraction of the vortex ring, a mechanism similar to the stabilisation of line vortices within a harmonically trapped BEC cloud~\cite{fetter:review}.

Examples of stationary states resulting from the imaginary time evolution are shown in \fref{2dpicture}. In a harmonic trap with $\omega_{r}=7.8 \times 2\pi$Hz and  $\omega_{z}=0.5\times 2\pi$Hz the BEC contains $2.2 \times 10^{6}$ atoms. The subplots are for two different sets of optical vortex parameters and ring flow charges: case~(i) and case~(ii), as given in the caption. Case~(ii) represents the more intense optical vortex. It can be seen in \fref{2dpicture}(a) that the single charge 7 ring singularity that was present in the initial state has broken up into a regular stack of seven unit charge ring singularities. For case~(i) they are responsible for the multi-peak structure of the Mach number shown in \fref{2dpicture}(c). If the parameters are altered towards those of case~(ii), the influence of the singularities on the flow in the tunnel is reduced. This corresponds to the transition from the solid to the dashed lines in \fref{2dpicture}(b-c). 

The integral along the $z$ axis of the flow velocity is directly connected to the ring vortex charge $q$. Due to the quantisation of circulation in a BEC, its velocity must obey 
$\oint_{\cal C}~\mathbf{v}~\cdot~\mathbf{dl}~=~\frac{h}{m}q$,
where $\cal C$ denotes any closed contour threading through all the ring singularities. As the BEC has a much smaller cross-sectional area available when it flows through the tunnel than when it returns on the outer shell of the cloud, the velocity in the tunnel is much higher. The above equation thus approximately becomes:
\begin{eqnarray}
\int_{-L/2}^{L/2} v_{z}(r=0,z)  dz&=\frac{h}{m}q,
\label{flowquant2}
\end{eqnarray}
where $L$ denotes the length of the optical tunnel in the $z$ direction. 

The presented stability analysis of stationary states is not completely general, as we impose cylindrical symmetry throughout the imaginary time evolution. An important class of ring vortex excitations, the Kelvin modes (see Refs.~\cite{dalibard:kelvins,fetter:kelvins} and references therein), could in principle break this cylindrical symmetry. However, a full three-dimensional (3D) stability analysis of skyrmions \cite{wuester:skyrm} indicates that the 2D model should give a very good indication of the stability properties:
In \rref{wuester:skyrm}, we have not found an indication of energetically unstable Kelvin modes of skyrmion's ring singularities. Nonetheless we point out that 3D studies of the optical tunnel would be an interesting, numerically challenging extension of our present work.

\subsection{Experimental creation scheme \label{experimental}}

The method of stabilising the ring singularities by pinning, proposed in the present article, should be experimentally realisable with present technology. The procedure to form the ring flow consists of initially creating a BEC at rest, already in the presence of the optical vortex. Then a ring flow structure with high vorticity is seeded by means of phase imprinting. In Ref.~\cite{ruoste:imprint2} J.~Ruostekoski and J.~R.~Anglin show in detail how the phase structure of the vortex ring, similar to the one given by Eq.~\bref{seedingstate_phase}, can be created. Their method uses a superposition of coherent light fields. In the presence of dissipation the BEC will, after the seeding, evolve towards stationary states like those in \fref{2dpicture}.

Compared to the previously known method to stabilise the ring vortex as a part of a skyrmion, our proposal is more flexible. A skyrmion is a topological soliton in a two component BEC, composed of ring vortex in one hyperfine component, which is filled with a second hyperfine component carrying a line vortex. This filling prevents the contraction of the ring vortex in the skyrmion.~\cite{savage:skyrm,wuester:skyrm}. However, even the stabilisation of a charge two ring flow in a skyrmion requires large atom numbers of the order $1 \times 10^{7}$, and more for higher charges~\cite{ruoste:highw}. In comparison, case (ii) represents a charge 9 ring flow in a condensate of only $2.2 \times 10^{6}$ atoms. 

\subsection{Supersonic flow\label{supersonic}}
Supersonic flow is expected to be energetically unstable~\cite{landau:superfluidity}. This manifests itself in the imaginary time evolution as a decay of the total ring flow charge from $q_1$ to $q_{2}<q_{1}$, whenever the parameters are such that a final state with charge $q_{1}$ would breach the speed of sound.
Despite the energetic instability, it is not generally true that the supersonic flow is also dynamically unstable~\cite{jackson:gpestab}.
Previously, it has been shown that dynamically stable flows with a supersonic region exist in a 1D ring system~\cite{garay:prl,garay:pra}, and that they can be obtained from the subsonic ones by an adiabatic change of trapping potential parameters.

Here we use the following sequence for the creation of supersonic flows: Initially the BEC is prepared in the stable state with subsonic ring flow corresponding to case~(ii). Then an additional potential as in Eq.~\eref{Vhump} is added dynamically. 
We found that for all potential widths, locations and ramp-up times employed, the generic response of the BEC was dynamical instability and the emission of a single or multiple grey solitons, travelling in the narrow confines of the optical tunnel. Refs. \cite{hakim:flow,law:hump} report on similar results in 1D simulations.
For our 2D simulations (that pertain to a 3D situation) we have employed an adaptive Runge-Kutta-Fehlberg method
within the high-level programming language XMDS~\cite{xmds_numerics}.

According to hydrodynamic theory, a hump potential with a single peak does not allow a quasi-stationary flow that is subsonic in a finite interval and subsonic on either end of this interval \cite{book:landlif,barcelo:towardshr}. Such a flow requires the realisation of a double de-Laval nozzle, using two hump potentials. The first of these renders the flow supersonic, while the second decelerates it back to subsonic velocities. We have numerically examined a variety of \emph{dynamical} sequences, inserting double de-Laval potentials into the existing ring flows, and found that the result was always the same: The speed of sound is breached initially at the hump farthest in the direction of BEC flow, thus not realising a double de-Laval structure. Nonetheless, we can not exclude that with more careful adjustment of the potential the stationary ring flows presented here could be used as a staging point for double de-Laval nozzles. These have been suggested for the creation of sonic horizons in the context of analogue gravity~\cite{barcelo:towardshr,sakagami:hr}.

\begin{figure}
\centering
\epsfig{file={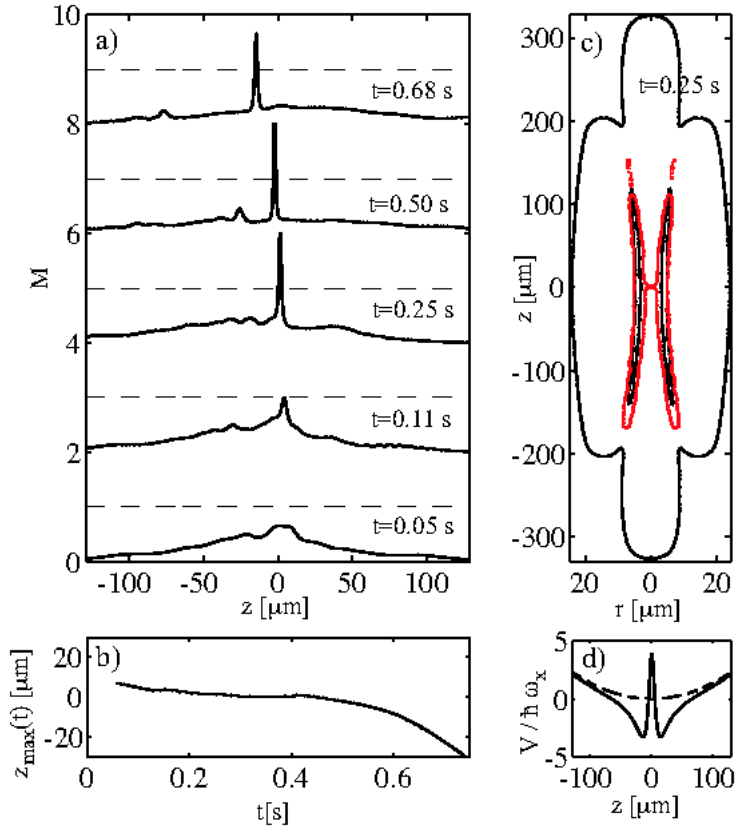},width=10cm} 
\caption{Scenario of long-time immobile soliton as sonic horizon, following a sudden change in the potential as shown in (d). (a) Time evolution of the Mach number $M$. Subsequent samples shifted by $2$. Dashed horizontal lines indicate $M=1$ for each time sample. The peaks of the Mach number are truncated at $M=2$. (b) Position of the maximum of $M$. (c) (red) Shape of the ergoregion~\cite{visser:analogue} and (black) the BEC cloud (shown is an isocontour at $\rho~=~3~\times~10^{16}~\mbox{m}^{-3})$. (d) The potential is suddenly modified (at $t=0$) from the shape indicated by the dashed line to the solid line.}
\label{localised_spike}
\end{figure}

We now present our results achieved with the dynamical addition of a particularly shaped single peak potential. For the \emph{sudden} addition of a hump potential, shaped as in \fref{localised_spike}(d), we find the following:
Once the flow exceeds the speed of sound a single grey soliton is emitted, and for time spans as long as $0.58$s it remains located near $z=0$, see \fref{localised_spike}. Only after that time does the grey soliton begin to accelerate. 
The potential was obtained using Eq.~\eref{Vhump} with parameters $V_{h1}=-8.5\hbar\omega_{r}$, $V_{h2}=12.5\hbar\omega_{r}$, $w_{1}=1.2\mu$m and $w_{2}=0.7\mu$m.

We observe that at all times the flow speed in the tunnel is a function of $z$ only, i.e.~it remains quasi one dimensional. It is thus justified to apply 1D soliton theory. We find excellent agreement between the functional form for 1D grey solitons \cite{theocharis:lagrangian,yuri:lagrangian} and the structures generated dynamically in our simulations. Further the dynamical behaviour of the soliton after its creation can be well understood with the aid of a 1D Lagrangian approach~\cite{theocharis:lagrangian,yuri:lagrangian}.  
To this end, we apply the Ansatz~\cite{theocharis:lagrangian,yuri:lagrangian}
\begin{eqnarray}
\psi(z,t)&=\sqrt{\sub{\rho}{bg}(z)} e^{-i \mu t/\hbar }u(z,t),
\label{lagrangian_ansatz}
\\
u(z,t)&=B(t) \tanh{[B(t)(z-z_0(t))/\xi(t)]} + iA(t).
\label{soliton_wavefunction}
\end{eqnarray}
Here $\mu$ is the chemical potential, $u(z,t)$ the soliton wavefunction and $\sub{\rho}{bg}(z)$ denotes the background condensate density. We work in a frame of reference that moves with the speed of the background flow ($\sub{v}{bg}$) at the soliton centre ($z_{0}$). This speed is allowed to vary slowly in time. The parameters in the wavefunction $A(t)$ and $B(t)$ are related to the solitons width and depth and $\xi(t)$ denotes the codensates healing length at $z_0(t)$. Further we have $A(t)+B(t)=1$. The leading order result for the soliton equation of motion following~\cite{theocharis:lagrangian,yuri:lagrangian} is:
\begin{eqnarray}
 \frac{\partial^{2} z_{o}}{\partial t^{2}}&=\frac{U}{4m}\left.\left(1 + \frac{U}{\mu}\sub{\rho}{bg}\right)\frac{\partial \sub{\rho}{bg}}{\partial z}\right|_{z=z_o}
 + \frac{\partial  \sub{v}{bg}}{\partial t}.
 \label{soliton_eom}
\end{eqnarray} 
We find that this equation describes the trajectories of solitons in our simulations well, within uncertainties arising from the numerical extraction of $\sub{\rho}{bg}$ and $\sub{v}{bg}$. In particular, Eq.~\eref{soliton_eom} explains the immobility of the soliton shown in \fref{localised_spike}(a-b). In the vicinity of the soliton's initial position the effective potential for the solitons, given by the background density, is approximately flat. Equation \bref{soliton_eom} suggests that the final travel direction of the soliton can be manipulated between positive and negative $z$ by using potentials which are slightly asymmetric along that axis, which we have confirmed numerically. The potential shape in such a case is still similar to \fref{localised_spike} (d), but the wells adjacent to the local maximum are of different depths. The BEC density increases on the side with the deeper well, making it the favoured escape direction for the grey soliton. 

%

\subsection{Prospects of analogue gravity \label{analogue}}

The purpose of the present section is to consider whether and how grey solitons could play a role in studies of analogue gravity. Our goal here is not to give a rigorous and complete account, but to present basic ideas and stimulate further interest.
 
The grey solitons obtained in \sref{supersonic} possess a sonic horizon where the BEC flow past them becomes supersonic, termed black hole horizon (BH). They also carry a white hole horizon (WH), the location where the flow returns to subsonic velocities once it has passed the soliton. The separation $d$ between the black-hole and while-hole horizons depends on the soliton size, which is determined by the healing length. We numerically find that $d\sim2\xi$. The conventional derivation of analogue Hawking radiation~\cite{visser:review} requires the length scale of variations in the BEC flow, and hence the separation between BH and WH, to be much greater than the condensate healing length. The reason for this is that only modes with phononic wave numbers $k$ such that $k\xi \gg 1$ propagate in direct analogy to scalar fields in curved space-time~\cite{unruh:bholes,visser:review,garay:prl,barcelo:diffmetric}.
\footnote{This issue is however a subtle one: All excitations experience a formally infinite blueshift i.e.~increase in wave number, when traced back in time towards the horizon. They thus necessarily enter the non-phononic part of the Bogoliubov spectrum. 
However, it was argued in Refs.~\cite{corley:hfdisp, leonhardt:bogolhawking} that the analogue Hawking effect persists despite this fact.}.

There exist interesting phenomena in analogue gravity, which crucially involve white holes in addition to black holes (e.g.~the black hole laser effect~\cite{corley:bhlaser}). However,  
BEC flow configurations with a widely separated black- and white-hole pair are usually dynamically unstable \cite{garay:prl,garay:bhstab}, resulting in grey soliton emission from the white-hole~\cite{garay:pra}. This is also the origin of the solitons discussed in \sref{supersonic}. As a grey soliton already represents the end-product of dynamical instability, the black and white holes contained \emph{within} it are stable, for as long as the soliton exists. 
Given the instabilities associated with well separated BH-WH pairs, compared to the ease with which one can obtain grey solitons, it would be interesting to enquire, which features of the Hawking or black-hole laser effects persist in a rapidly varying flow. Further motivation for research on quantum emission properties of rapidly varying condensate flows arises from the strict limits on the achievable analogue Hawking temperatures as long as the flow varies slow enough to be considered in the hydrodynamic regime~\cite{wuester:horizons}. It was shown that these arise due to three-body losses.

It has been stated that the core ingredients of Hawking radiation are exclusively the existence of an apparent horizon and the validity of the eikonal approximation there~\cite{visser:essential}. 
The Mach number around the long time immobile soliton becomes as high as $600$\footnote{The Mach number is truncated at $M=2$ in \frefp{localised_spike}{a}.}. The group velocity $v_{g}$ of Bogoliubov excitations with wave number $k$ is given by~\cite{barcelo:diffmetric}:
\begin{eqnarray}
|v_{g}|&=\frac{c^{2}+\frac{\hbar^{2}}{2m^{2}} k^{2} }{\sqrt{c^{2}k^{2} + \left(\frac{\hbar}{2m}k^{2} \right)^{2}}}|k|.
\label{group_velocity}
\end{eqnarray}

Inserting the speed of sound near the soliton, we see that only wave packets comprised of wavelengths shorter than $g\xi$ with $g=0.015$ have a group velocity in excess of $600c$. There is thus a window of wave numbers, $\pi/\xi \ll k \ll  2\pi/(g\xi)$, for which a point of no return exists \emph{and} the eikonal approximation should be valid in its vicinity.
The location of this ``horizon'' varies depending on the major wavelength of the wave packet. Due to the steep Mach number profile, the ``horizons'' for different wavelengths are spatially close. Quantum emission of quasi-particles thus appears conceivable for wave numbers within the window, however further research is required to determine the existence of an effect.  

Another interesting analogue gravity system, for which a BH-WH pair associated with the propagation of a soliton was previously discussed is fermionic superfluid $^{3}$He-A~\cite{volovik:solitonhorizon}. There the soliton of choice is a domain wall between regions of parallel and anti-parallel orientation of the spin and orbital angular momentum of Cooper-pairs. In such a system the separation between BH and WH can significantly exceed the coherence length \cite{volovik:review}. Owing to the structure of the superfluid order parameter near the domain wall, the scenario discussed in \rref{volovik:solitonhorizon} realises a rotating analogue BH, with an ergoregion boundary distinct from the apparent horizon \cite{volovik:solitonhorizon}, in contrast to the simpler setup under consideration here.
 
Returning to our condensate scenario, the velocity of the bulk BEC beyond the tunnel is effectively zero, see \frefp{2dpicture}{c}. This provides a region of flat space for the detection of quasi-particles emitted by the horizon. The steering of the soliton's escape direction could be an advantage for spectroscopic detection of horizon radiation \cite{schuetzhold:phonondetection}. The horizon-soliton can be made to travel towards $z>0$, while the generated quantum excitations might be spectroscopically detected in the bulk BEC at $z<0$ before the soliton reaches the cloud edge. 

One of the advantages of the analogue gravity program over the astrophysical original, is the ability to push the studied system to the validity limits of existing derivations of horizon radiation. This allows the segregation of essential premises from merely analytically convenient assumptions. In this sense our system might one day complement more conventional analogue gravity setups along the lines proposed in Refs.~\cite{garay:prl,giovanazzi:horizon,barcelo:towardshr,sakagami:hr,garay:pra}. 

\section{Conclusions\label{conclusions}}

We have introduced a novel arrangement for the studies of persistent ring flows in a BEC. Our method utilises optical vortices, laser beams with a phase singularity at the centre and hence a zero intensity hollow core. If tightly focussed, the highest intensity region of the optical vortex can fit into the low density region of the matter-wave ring vortex in the BEC, stabilising the ring singularity against contraction. We numerically showed that energetically stable BEC states with ring flows of charges up to $q=9$ exist in our system. 

We have further analysed the response of the ring flows to dynamically varied potentials which drive the flow into the supersonic regime, realising a supersonic optical tunnel for the BEC. In this case, our simulations show the creation of quasi one dimensional grey solitons in the tunnel. The solitons appear due to the dynamical instability once the speed of sound is reached by the velocity of the BEC flow. A single grey soliton that is immobile in the lab-frame for a long time was found for a particular choice of potential.

Finally, we have discussed the prospects of using this single, stabilised soliton as a tool for analogue gravity.  We point out that whilst the Mach number around the realised horizon varies too rapidly to expect the usual form of analogue Hawking radiation, the soliton's sonic horizons might be suited to probe the physical boundaries for quantum emission from a horizon. 

\ack{We gratefully thank C.~Savage and E.~Ostrovskaya for their help. It is also a pleasure to thank A.~Bradley, J.~Close, R.~Fischer, J.~Hope, P.~Jain, D.~Neshev, A.~Truscott and W. Kr\'olikowski for useful discussions. This research was supported by the Australian Research Council under the Centre of Excellence for Quantum-Atom Optics and by an award under the Merit Allocation Scheme of the National Facility of the Australian Partnership for Advanced Computing.}

\end{document}